\documentclass[12pt,preprint]{aastex}

\usepackage{amsmath}
\usepackage{graphicx}
\usepackage{multirow}
\usepackage{color}
\usepackage{natbib}
\usepackage{enumerate}
\font\sm=cmss10 at 10truept

\def\A{{\cal A}}
\def\C{{\cal C}}
\def\S{{\cal S}}

\def\Hpert{{{\cal H}_{\mathrm{pert}}}}

\begin{document}

\title{Corralling a distant planet \\ with extreme resonant Kuiper belt objects}

\author{Renu Malhotra\footnote{Please address correspondence to: renu@lpl.arizona.edu} \\Kathryn Volk \\
Xianyu Wang\footnote{Also: School of Aerospace Engineering, Tsinghua University, Beijing, China}}
\affil{Lunar and Planetary Laboratory, The University of Arizona, Tucson, AZ 85721, USA.}


\begin{abstract}
The four longest period Kuiper belt objects have orbital periods close to integer ratios
with each other.  A hypothetical planet with orbital period $\sim$~17,117~years, semimajor axis 
$\sim$~665~AU, would have N/1 and N/2 period ratios with these four objects.  
The orbital geometries and dynamics of resonant orbits constrain 
the orbital plane, the orbital eccentricity and the mass of such a planet, as well as its current
location in its orbital path.
\end{abstract}


\section{The most distant Kuiper belt objects}

In the outer solar system beyond the orbit of Neptune is a large belt of minor planets, the Kuiper belt, whose dynamics has been understood, over the past two decades, to be significantly controlled by the gravity of the giant planets, Jupiter---Neptune, either through secular or resonant perturbations or by gravitational scattering.
However, there are a few known Kuiper belt objects [KBOs] that are unlikely to be significantly perturbed by the known giant planets in their current orbits.  
In the dynamical classification scheme for the small bodies of the outer solar system, such objects (semimajor axis, $a$, and perihelion distance, $q$, exceeding 150 AU and 40 AU, respectively) would likely belong in a class known as ``detached objects'', distinct from the classical KBOs, the resonant KBOs and the scattered/scattering KBOs~\citep{Gladman:2008}. 
\cite{Trujillo:2014} call them ``extreme'' KBOs [eKBOs, henceforth].   
Upon querying the Minor Planet Center's catalog, we identify six KBOs having $a>150$~AU and $q>40$~AU; two of these are numbered and four are unnumbered: 
Sedna (90377), 2010 GB174, 2004 VN112, 2012 VP113, (148209), 2013 GP136.
Their orbital inclinations are in the range 11.9--33.5 degrees (to the ecliptic), not especially remarkable for KBOs. However all of these objects have extremely large orbital eccentricity, $e\gtrsim0.7$, suggesting large orbital perturbation in their past history.  
This is rather at odds with their current large separations from the known giant planets.

A natural hypothesis is that these objects' orbital dynamics were or are dominated by the gravitational perturbations of a 
planet in the heliocentric distance range of a few hundred AU.  \cite{Gladman:2006} hypothesized a ``rogue'' planet, a short-lived
planet larger than an earth-mass, to explain the eKBOs. 
\cite{Trujillo:2014} proposed an extant hypothetical distant planet to explain an apparent clustering of the arguments of 
perihelion of the eKBOs. 
\cite{Batygin:2016} have also argued for this hypothesis by calling attention to the clustering of the 
eccentricity vectors and orbit poles of some distant KBOs.  These authors suggest a planet mass of $\sim 10M_\oplus$, and an orbit with semimajor axis $a\simeq 700$~AU, and eccentricity $e\simeq0.6$. 
\cite{Fienga:2016} provide an analysis of {\it Cassini} tracking data, arguing that, for the favored planet parameters of \cite{Batygin:2016}, the residuals in the orbital motion of Saturn are reduced with the existence of such a planet at current true longitude near $120^\circ$; however their analysis cannot exclude a large range of true longitudes, $(-132^\circ,106^\circ)$, where the planet would be too distant to sufficiently perturb Saturn. 

Here we report additional support for the extant distant planet hypothesis, and we narrow the range of its 
parameters and its current location.  
We point out hitherto unnoticed peculiarities of the orbits of the eKBOs:  
we find that the orbital period ratios of these objects are close to integer ratios. 
Such ratios are not dynamically significant unless the eKBOs are in mean motion resonances (MMRs) with a massive planet. 
We identify an orbital period of the hypothetical planet which is in simple integer ratios 
with the four most distant eKBOs.  We show that the resonant dynamics provide constraints on the 
planet's orbital parameters and its mass and can also narrow the possible range of its current location 
in its orbital path.

\section{Mean motion resonances}

The wide range of the perihelion--to-aphelion distances of each eKBO may overlap 
those of the hypothetical planet [HP, henceforth].   In this circumstance, the extreme eccentricities of the eKBOs may
be a natural consequence of gravitational scattering encounters with the planet. 
The dynamical survival of scattered objects over solar system age could be simply due to the long timescales of 
close encounters at these distances, or it could be due to enhanced dynamical stability offered by phase-protection 
in mean motion resonances with HP.  
In their analysis, \cite{Batygin:2016} emphasized the secular effects of the planets in aligning the orbits of the eKBOs, but also noted that transient resonant interactions would be important.
Here we investigate the possibility of extant MMRs of the known eKBOs with the unseen planet.

Examining the orbital periods of the six eKBOs as listed in the Minor Planet Center's
catalog, and ordering them in order of decreasing orbital period 
(so Sedna is eKBO\#1, with the longest period orbit), 
here is what we find for the ratios $P_1/P_j$: 
$$ 1.596, 1.993, 2.666, 3.303, 6.115.$$  
These values are close to the following rational ratios: 8/5, 2/1, 8/3, 10/3, and 6/1, sufficiently intriguing to investigate further.

We proceeded as follows.
We computed the best-fit orbits for these six eKBOs, as well as their uncertainties, 
using the orbit-fitting software developed by \cite{Bernstein:2000}; 
for these orbit-fits, we included all of the astrometric measurements for the objects listed in the Minor Planet Center.
We find the following best-fit semimajor axes and their 1--$\sigma$ uncertainties (all quoted in AU):
$$506.84\pm0.51, 350.7\pm4.7, 319.6\pm6.0, 265.8\pm3.3,221.59\pm0.16,149.84\pm0.47.$$
These osculating barycentric elements agree well with 
the heliocentric MPC orbits when both are converted to heliocentric cartesian positions and velocities.

We then consider some simple possibilities for HP's resonant orbit. 
Assume that its orbital period $P'$ is longer than $P_1$, but not by too much, else its gravitational effects would 
not be very strong.  
So, we consider $P'/P_1$ values of 2/1 or 3/2 or 4/3.
We find that the case of $P'/P_1 = 3/2$ (hence $a'\approx665$~AU) is particularly striking, for it yields period ratios close to 
5/2, 3/1, and 4/1 for $P_2$, $P_3$ and $P_4$. 
The other simple choices ($P'/P_1$ values of 2/1 or 4/3) yield larger integer ratios for $P'/P_2, P'/P_3,P'/P_4$.  Although these larger integer ratios cannot be ruled out {\it prima facie}, we limit this first study to the small integer ratio MMRs; other MMRs will be considered in a future study.

\begin{figure}[h]
\center
\includegraphics[width=4.2truein,angle=270]{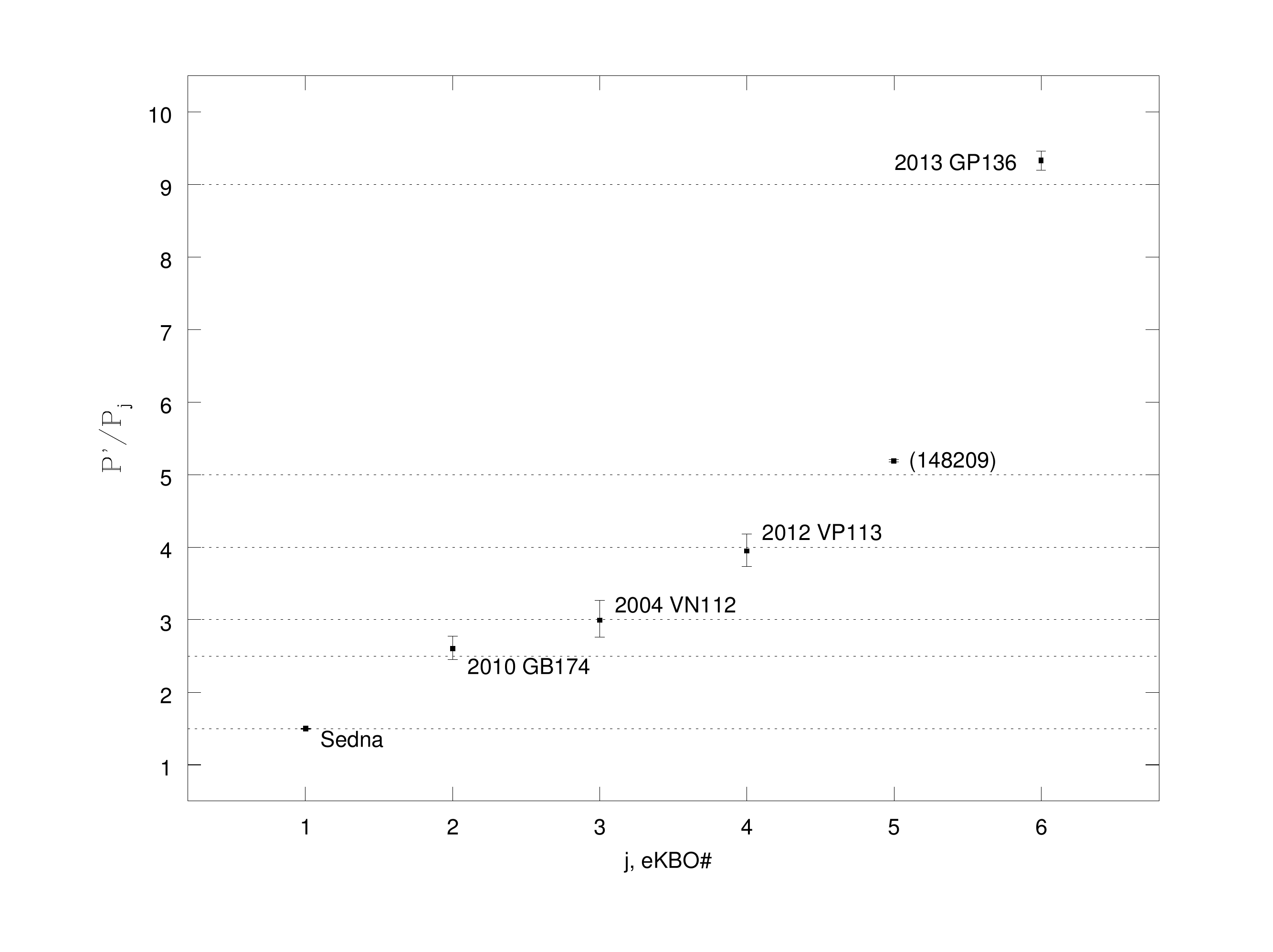}
\caption{\sm
Period ratios of the six eKBOs with the hypothetical planet.  The errorbars indicate 3--$\sigma$ uncertainties.   The orbit period $P'$ of HP is assumed to be 3/2 times Sedna's best-fit orbital period, $P_1=$~11,411~years.  The dotted horizontal lines indicate the rational ratios 3/2, 5/2, 3/1 4/1, 5/1 and 9/1.}
\label{f:pratios}
\end{figure}

In Fig.~\ref{f:pratios}, we plot the period ratios $P'/P_j$ for the best-fit orbits with their uncertainties. 
We observe that, within 3--$\sigma$ uncertainties, four of the six eKBOs have periods near N/1 or N/2 ratios with HP, 
when we take the latter's period to be 1.5 times Sedna's best-fit period.  
We also observe that the error bars are not insubstantial and one can be skeptical of any perceived coincidence with 
simple ratios.  
We remark that our procedure for computing the uncertainties weights all published observations equally and may overestimate the true errors.  

Secondly, one might also be skeptical of the period ratio coincidences because not all possible N/1 and N/2 period 
ratios in the range shown are occupied.  
On this point, we can remark that (a) the larger N values lead to shorter period orbits in which HP's influence is
diminished and Neptune's influence looms larger, potentially destabilizing the resonances with HP, and 
(b) the small number of currently known eKBOs will occupy only a small fraction of all possible N/1, N/2 period ratios. 

Thirdly, one can ask if the proximity to the resonant ratios is different from random.  
We answer this question approximately, as follows\footnote{A rigorous statistical analysis finds similar results (Sean Mills \& Daniel Fabrycky, personal communication).}.  
For integer values $0<N_i<N_j\leq10$, there are 31 distinct rational numbers $N_j/N_i$ in the range $(1,10)$, so the mean distance between said rationals is $\sim0.29$.  
We can compare this mean value to the distances of the observed period ratios to the resonant ratios. 
Taking account of the 3--$\sigma$ observational uncertainties, the latter are 0.007, 0.27, 0.27, 0.26, 0.21, and 0.46, respectively (listed in increasing period ratio, as in Fig.~\ref{f:pratios}). 
The shortest period eKBO, 2013 GP136, presents the largest distance (from the 9/1 ratio), as is evident in Fig.~\ref{f:pratios}.
Sedna's very small distance to the 3/2 ratio is explained by its very small orbital uncertainties and because its period ratio of 3/2 is fixed by hypothesis. 
The remaining four values are similar to the mean distance between the rationals, indicating that they are not strongly distinct from random.  
However, this statistical argument depends upon how many rationals are allowed and it does not consider any physics of MMRs. 
We will show later (Section 2.2) that the libration widths of the specific MMRs are similar to these apparent distances from exact resonant ratios; 
hence, libration in MMRs is possible even with the significant apparent distances from exact resonant ratios.
We therefore proceed with investigating the implications of these MMRs for the four longest period eKBOs.

\begin{figure}[h]
\centering
\vspace{-0.4truein}
\hglue1truein\includegraphics[width=6.4truein]{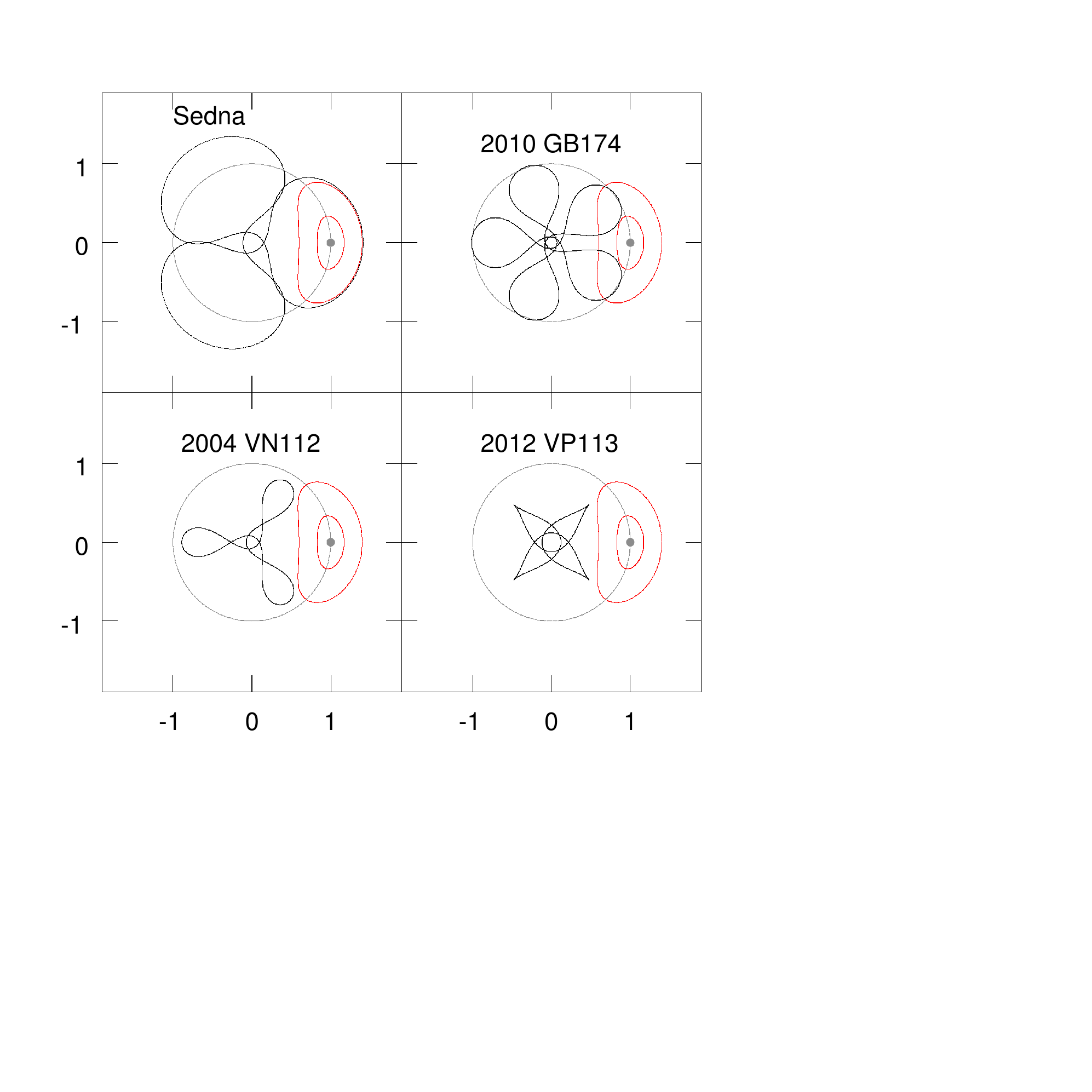}
\vglue-2.1truein
\caption{\sm
Resonant orbits in the rotating frame are in black.   The red loops indicate the trace of HP in possible eccentric orbits of eccentricity 0.17 and 0.4.   
Distances are scaled to the semimajor axis of HP.  The grey circle is of unit radius.  }  
\label{f:resonantorbits}
\end{figure}


\subsection{Resonant orbital geometry constraints}

Stable resonant orbits of the eKBOs will avoid close approaches with HP, and will have a discrete number of 
geometries for the conjunctions of the eKBO with the planet.
These are described by librations of the critical resonant angle, 
\begin{equation}
\phi =(p+q)\lambda' - p\lambda -q\varpi.
\end{equation}
As usual, in the definition of the resonant angle, $p$ and $q$ are integers, 
$\lambda$ the mean longitudes and $\varpi$ the longitudes of perihelion; we use
the prime to refer to orbital parameters of the planet, the parameters of the eKBO are unprimed. 
Depending upon the orbital eccentricity, $\phi$ may be stable at either 0 or at $\pi$ or at both.

We illustrate in Fig.~\ref{f:resonantorbits} the trace of the resonant orbits (of zero libration amplitude) 
in the frame co-rotating with the mean angular velocity of HP.   
Also included here are the traces of a few possible coplanar orbits of HP, 
with eccentricity $e'=0, 0.17$ and $0.4$.  

We observe that for Sedna, stable libration is likely only near $\phi=\pi$,
and unlikely near $\phi=0$, because the latter would allow small encounter 
distances.  
(This is the opposite of the geometry for the low eccentricity regime
in which stable (unstable) librations are centered at $\phi=0(\pi)$.)
Consequently, conjunctions with HP occur near one of three longitudes, 
$\pm\pi/3$ and $\pi$ from Sedna's longitude of perihelion.
Moreover, the long term stability of Sedna would limit $e'\lesssim0.41$.
Likewise, a coplanar orbit of 2010 GB174 with $\phi=\pi$ 
limits $e' <0.04$; however, in the geometry with $\phi=0$, 
the limit is $e'\lesssim0.18$.  
The orbit of 2004 VN112 limits $e'\lesssim0.42$ for $\phi=\pi$. 

\begin{figure}[h]
\centering
\vglue-1.2truein\includegraphics[scale=0.6,angle=270]{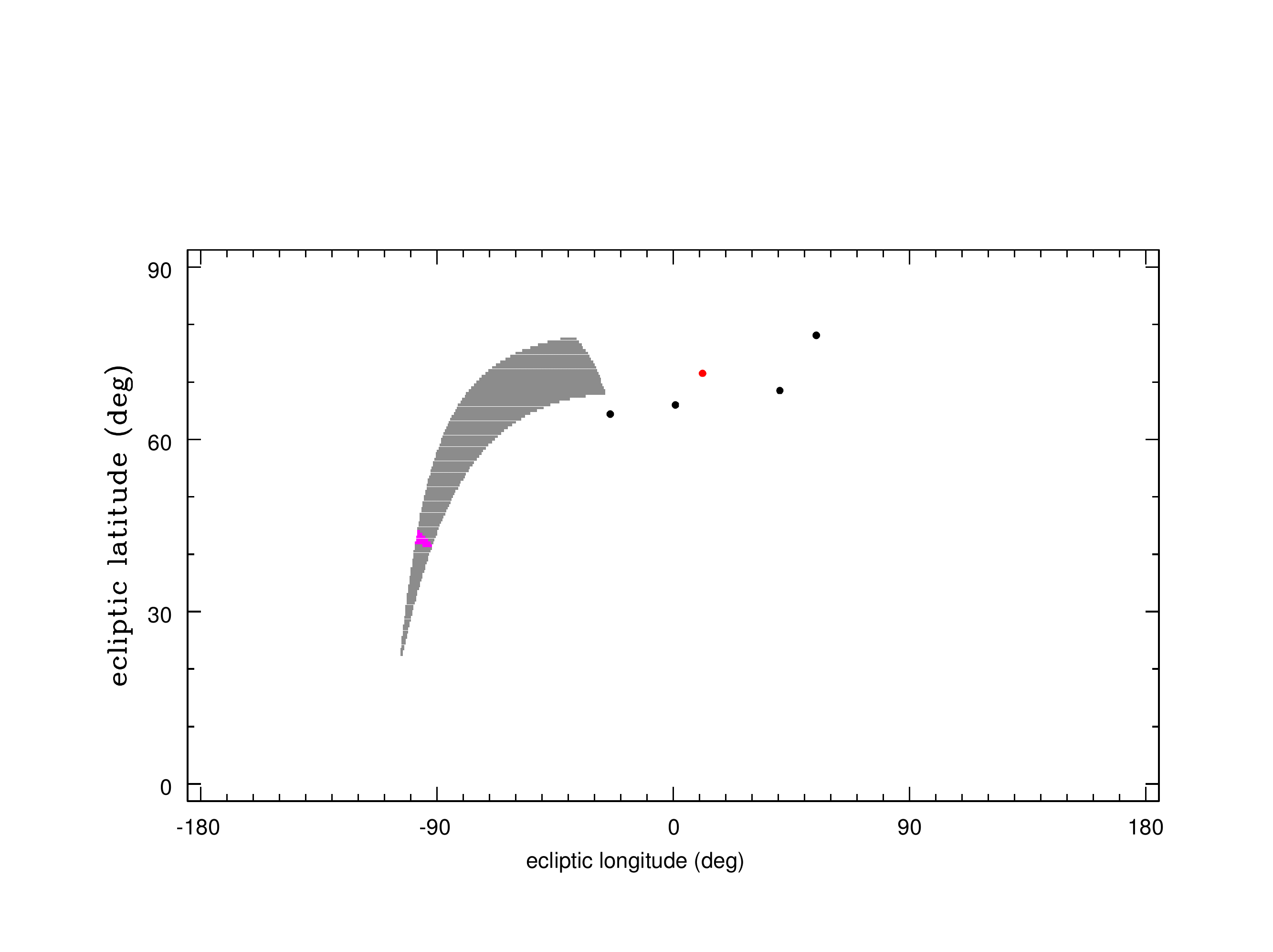}
\vglue-0.2truein\caption{\sm
The orbit planes, as described by the ecliptic latitude and longitude of the orbital normal vectors, of the four
potentially resonant
eKBOs are indicated by the black dots.  The red dot indicates their mean plane.  The region shaded in grey identifies the planes for which the four eKBOs have $g$ within 45 degrees of $\pm\pi/2$.  The region shaded in magenta identifies the subset for which the relative inclinations of the eKBOs are also within 10 degrees of the stationary inclination values at the periodic orbit of the third kind.}  
\label{f:hpplane}
\end{figure}

Of course, strict coplanarity of HP is possible for at most one eKBO, because the eKBOs' 
orbits are not coplanar.  Fig.~\ref{f:hpplane} shows the ecliptic latitude and longitude of 
the orbit normal vectors of each of the four potentially resonant eKBOs.  
We see that they are somewhat clustered, approximately within one 
quadrant of ecliptic longitudes; they have relative inclinations 7--15 degrees to their mean plane.

The non-coplanarity of the orbits with HP's orbit plane may relax the upper limit on $e'$, especially if the eKBO's argument of perihelion
$g$ (relative to HP's orbit plane) were near $\pm90^\circ$. 
This would have the effect of lifting the eKBO's aphelion away from the HP orbit plane and thereby increasing 
the closest approach distances and enhancing dynamical stability\footnote{This geometry is reminiscent of the 
dynamics of Pluto's orbit, which librates in an exterior 3/2 mean motion resonance with Neptune; Pluto's 
argument of perihelion is presently $\sim104^\circ$, and it librates about $90^\circ$ with a period of about 
4 megayears~\citep{Malhotra:1997}.}.
Such a configuration is associated with the so-called ``periodic orbits of the third kind'' 
envisioned by Henri Poincar\'e (as noted by \cite{Jefferys:1966a}) and computed by 
\cite{Jefferys:1966b} \cite{Kozai:1969} and \cite{Jefferys:1972} for the three dimensional circular restricted 
three body problem.  Such periodic orbits are characterized by simultaneous stationarity 
of the resonant angle $\phi$ (at 0 or $\pi$) and of the argument of perihelion $g$ (at $0$ or $\pi$ or $\pm\pi/2$).  
Physically, we can visualize these periodic orbits as having a fixed orbit plane and a 
fixed eccentricity vector (i.e., the apsidal precession rates of both node and pericenter vanish).
In the particular case of $g=\pm\pi/2$, the particle reaches aphelion when it is at its largest excursion away from the plane of the 
planet. 

We computed the values of $g$ relative to the mean plane of the four longest period eKBOs, finding that they are 
$$145.4^\circ, -70.4^\circ, 12.0^\circ, -39.3^\circ.$$

We look for an alternative HP orbit plane for which $g$ is proximate to $\pm90^\circ$, within some specified
tolerance, for all of the four eKBOs. 
We proceed as follows.
Let $\hat{\bf n}$ be the (unknown) unit vector normal to the HP orbit plane.
Let $\hat{\bf c}_j$ be the unit vector in the orbit plane of an eKBO and normal to its eccentricity vector. 
We ask whether $\hat{\bf c}_j$ can be rotated by no more than a specified amount, $\Delta\omega$, in the eKBO's orbit plane, such that $\hat{\bf n}\cdot\hat{\bf c}_j$ changes sign; if so, this would indicate that the eKBO has $g$ within $\Delta\omega$ of $\pm\pi/2$, near the periodic orbit of the 3rd kind.  By demanding that all four eKBOs satisfy this constraint, we identify the range of possible $\hat{\bf n}$.   The results of this search are shown in Fig.~\ref{f:hpplane}, in which we plot in grey the ecliptic longitude and latitude of the allowed $\hat{\bf n}$'s for $\Delta\omega=45^\circ$.   

We narrow the possible HP planes further by examining the inclinations of the eKBOs to possible HP orbit planes. 
Consulting the results of \cite{Jefferys:1972}, we find that the stationary inclinations (relative to HP's orbit plane) are 
$49^\circ, 62^\circ, 54^\circ$, and $57^\circ$, respectively, for the specific MMRs and the observed eccentricities of the four eKBOs.  
Of course, these eKBOs are likely librating with non-zero amplitude about the pertinent periodic orbit, so we look for those 
$\hat{\bf n}$'s that yield inclinations within $10^\circ$ of these values.  
Carrying out this search, we find the range indicated in magenta in Fig.~\ref{f:hpplane} satisfies this constraint.  
We conclude that an HP orbit plane with inclination near $i\approx 48^\circ$ and longitude of ascending node near 
$\Omega\approx-5^\circ$ allows all four resonant eKBOs to be in proximity of the periodic orbits of the third kind.

\subsection{Planet mass}

Resonant perturbations from HP provide phase-protection if  
an eKBO is sufficiently close to the exact resonant orbit.  
How close is close enough depends upon the planet mass. We now quantify this condition. 

In the usual canonical variables, 
\begin{eqnarray}
J_1 = L &=& \sqrt{GM_\odot a}, \qquad \theta_1 = l\ \hbox{(mean anomaly)} \\
J_2 = G &=& L\sqrt{1-e^2}, \qquad \theta_2 = g\ \hbox{(argument of pericenter)},\\
J_3 = H &=& G\cos i, \quad\qquad \theta_3 = h\  \hbox{(longitude of ascending node)},
\end{eqnarray}
the orbital perturbation equations are given by
\begin{equation}
\dot J_i =-{\partial \Hpert\over\partial \theta_i}, \qquad 
\dot \theta_i = {\partial \Hpert\over\partial J_i}, 
\end{equation}
where the perturbation Hamiltonian is
\begin{equation}
\Hpert = -{Gm'}( {1\over |{\bf r}-{\bf r}'|} - {{\bf r}\cdot{\bf r}'\over r'^3})
 \simeq -{Gm'\over a'} \A(L,G,H) \C(\phi).
\end{equation}
Here the first equation gives the Newtonian perturbation potential (${\bf r}$, ${\bf r}'$ are the 
position vectors of an eKBO and of HP, respectively, in Jacobi coordinates~\citep[e.g.,][]{Brouwer:1961};
i.e., they are relative to the barycenter of the Sun and the known planets), 
whereas the second equation gives an approximation for the resonant perturbation. 
The dependence of the resonant perturbation on the canonical variables
can be expressed as a product of an amplitude $\A(L,G,H)$ 
which is a function of only the canonical momenta, 
and a phase $\C(\phi)$ which is an even function of the resonant angle.  
($\A$ also depends on the orbital parameters of HP,
which will be assumed to be constant.)
Without loss of generality, we can define $\A$ and $\C$ such that $|\C(\phi)| < 1$ for $0\leq\phi<2\pi$. 
For the eKBOs whose extreme eccentricities are near unity, the magnitude of $\A$ will be of order a few.
When we require numerical values (below), we will adopt $\A=3$; but this value can, in principle, 
be determined more accurately with numerical averaging of the perturbation potential.

The mean motion of the eKBO is $n=(GM_\odot)^2/L^3$. Therefore, the resonant perturbations of the mean motion
are given by $\dot n = -(3pn/L)\partial\Hpert/\partial\phi$, whence we obtain the following
equation of motion for $\phi$,
\begin{equation}
\ddot\phi \simeq - {3p^2 m' a\over M_\odot a'}n^2 \A~\S(\phi),
\end{equation}
where $\S(\phi) =-d\C(\phi)/d\phi$. 
This equation for $\phi$ resembles the nonlinear pendulum. The small amplitude
oscillation frequency is given by 
\begin{equation}
\omega_0 \simeq |{3p^2m' a\over M_\odot a'} \A|^{1\over2}n.
\end{equation}
For the resonances of interest, the corresponding libration periods are on the order of a few megayears 
for a few-earth-mass planet.

By analogy with the pendulum, we divine that $\phi$ (hence the conjunction longitudes $\lambda_c$) can execute 
librations about stable equilibrium points provided $|\dot\phi|=|p\delta n + q\dot\varpi| \lesssim 2\omega_0$.  
Thus the required condition for resonance is that the deviation of the mean motion of the eKBO from the 
exact resonant value, $(p+q)n'/p$, should not exceed $\sim2\omega_0/p$.  Then, by Kepler's third law, 
the resonance width in semimajor axis is given by
\begin{equation}
\Delta a_{\mathrm {res}} \simeq a|{16 m' a \A \over 3 M_\odot a'} |^{1\over2}
 \simeq 0.007a|{m'a \A \over 3 M_\oplus a'} |^{1\over2}.
\label{e:dares}\end{equation}
Computing the widths of the 3/2, 5/2, 3/1 and 4/1 resonances for the four eKBOs of interest, we find that
a $10M_\oplus$ planet yields $\Delta a_{\mathrm {res}}$ values comparable to the uncertainties
in the semimajor axes of these objects. 

The more severe condition on the planet mass obtains when we demand that the eKBOs are in libration about 
periodic orbits of the third kind.
The frequency of small amplitude oscillations about these periodic orbits\footnote{These periodic orbits,
embedded in the $N/1$ or $N/2$ MMRs, are often inaccurately referred to as ``Kozai within MMR''.  They 
can be considered to be the discrete sequence whose limit for $N\gg1$ is the 
Kozai-Lidov resonance~\citep{Kozai:1962,Lidov:1962}. Eq.~\ref{e:w0KLC} is obtained in this limit.} 
is given by
\begin{equation}
\omega_0\approx c\, n {m'a^3\over M_\odot b'^3}e\sin i,
\label{e:w0KLC}\end{equation}
where $n,a,e,i$ are a resonant eKBO's mean motion, semimajor axis, eccentricity and inclination (relative to HP's orbit plane), 
$b'=a'\sqrt{1-e'^2}$ is the planet's semi-minor axis, and $c$ is a numerical coefficient whose asymptotic value is $\sim 5.8$ 
for $a\ll a'$ ($c$ will be larger for $a$ approaching $a'$). For numerical estimates, we adopt $e'=0.2$ and $c=8$.
We find that, for $m'=10M_\oplus$, the small amplitude oscillation periods for the four eKBOs are 0.6--0.94~Gyr, a factor of a few
smaller than the age of the solar system.
We conclude that $10M_\oplus$ is approximately the minimum mass necessary to maintain periodic orbits of the third kind
for the four eKBOs.

\subsection{Planet's current location in its orbit}

Each resonant eKBO constrains the longitude of HP because the critical resonant angle must librate about either 0 or $180^\circ$. 
For illustration, consider the libration of 
$\phi=3\lambda'-2\lambda-\varpi$ about $180^\circ$ for Sedna's 3/2 resonant orbit.  
Given that Sedna is presently near its perihelion, $\lambda\approx\varpi$, the planet can be in proximity to one of 
three longitudes: $60^\circ$ or $180^\circ$ or $300^\circ$ from Sedna's longitude of perihelion, $\varpi$.  
The trace of the resonant orbit in the rotating frame (Fig.~\ref{f:resonantorbits}) shows that 
libration of $\phi$ would allow the planet to be located as far as $\sim54^\circ$ from these 
longitudes. 

This means that only three small zones of $\pm6^\circ$ around each of three locations, $\varpi$, $\varpi+120^\circ$ 
and $\varpi-120^\circ$ are excluded for the planet's present location.
For $e'=0.2$, we find that the excluded zone is a little larger, $\pm7.5^\circ$ around each of the 
same three locations.

\begin{figure}[h]
\centering
\begin{minipage}[t]{0.45\textwidth}
\includegraphics[width=\textwidth]{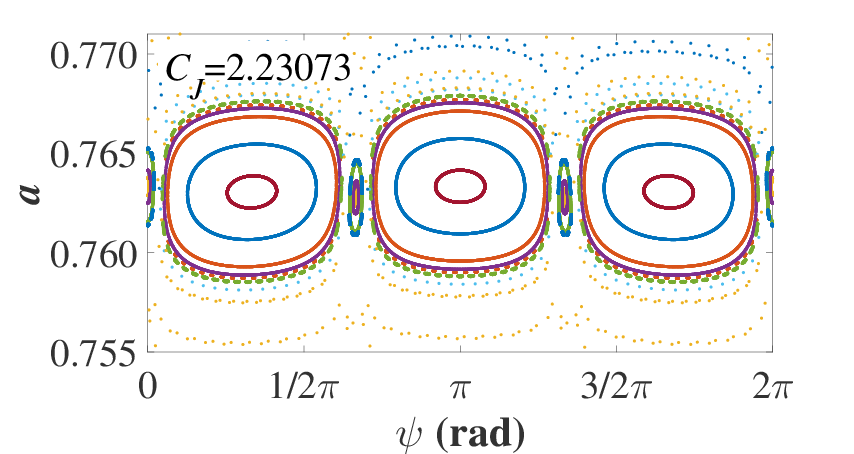}
\end{minipage}
\begin{minipage}[t]{0.45\textwidth}
\includegraphics[width=\textwidth]{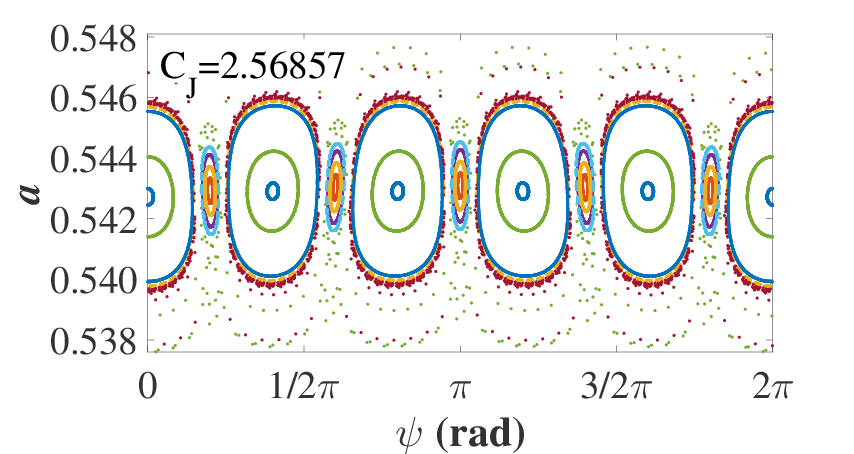}
\end{minipage}
\begin{minipage}[t]{0.45\textwidth}
\includegraphics[width=\textwidth]{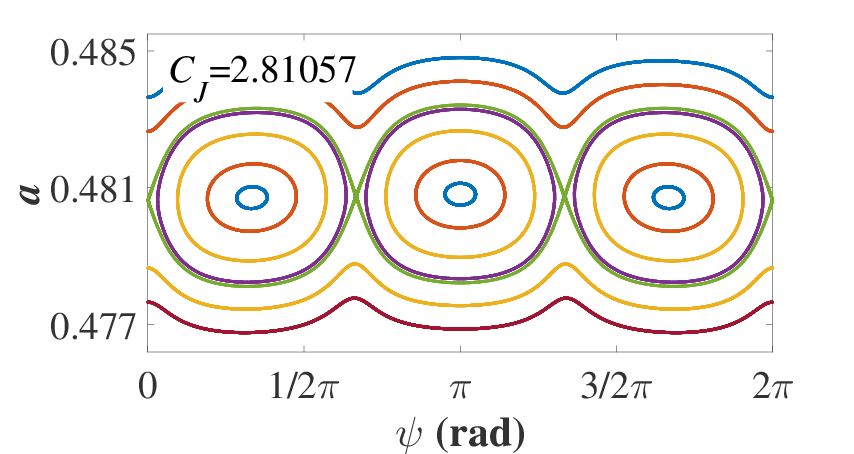}
\end{minipage}
\begin{minipage}[t]{0.45\textwidth}
\includegraphics[width=\textwidth]{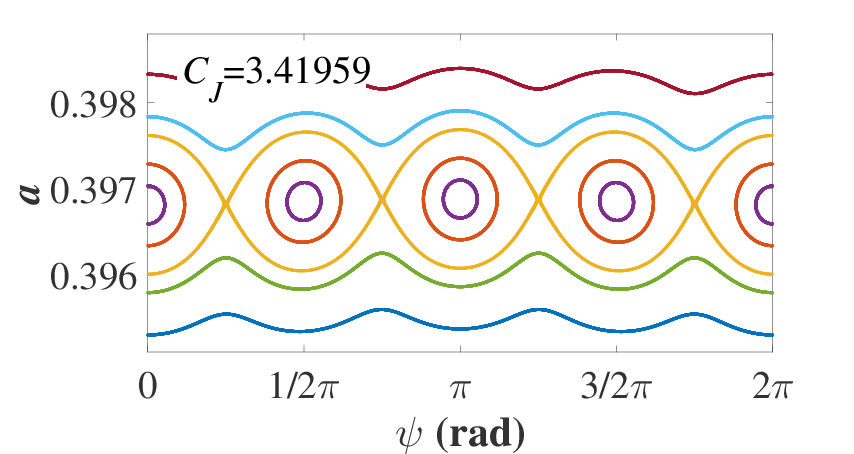}
\end{minipage}
\caption{\sm
Surfaces of section for the circular planar restricted three body problem, for mass ratio $m_2/(m_1+m_2)=3\times10^{-5}$,
and Jacobi constant values as indicated. These are phase space portraits in the vicinity of the 3/2, 5/2, 3/1 and 4/1 resonant
orbits of eccentricity 0.85, 0.87, 0.85 and 0.7, respectively.  The abscissa, $\psi$, measures the
angle between the planet and the particle's pericenter; the ordinate is the particle's barycentric semimajor axis, in units of the circular orbit radius of the planet.}
\label{f:sos}
\end{figure}

\begin{figure}[h]
\centering
\includegraphics[scale=0.6,angle=270]{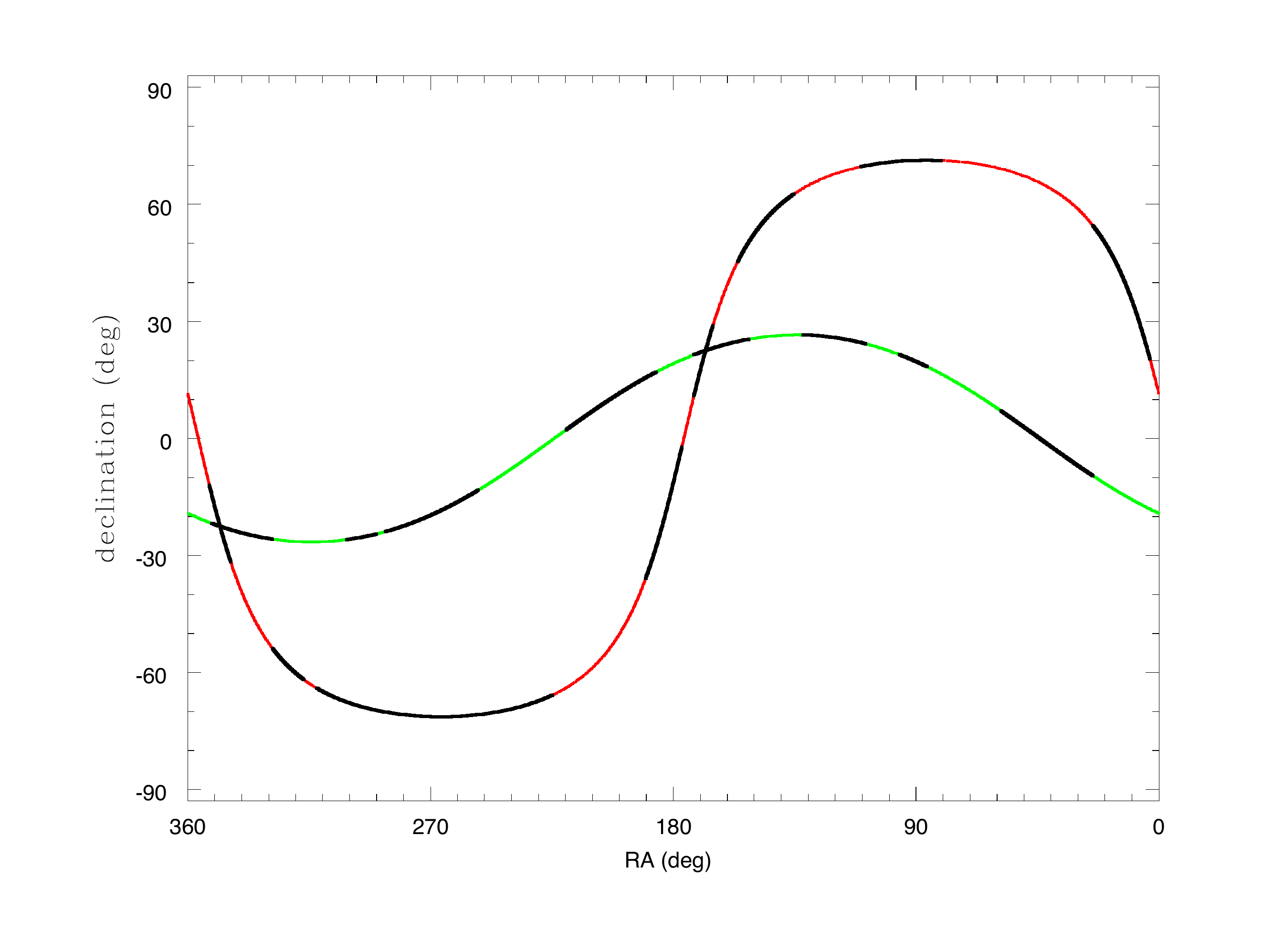}
\vspace{-0.3truein}
\caption{\sm
The possible locations of HP in the sky: in green for a representative low inclination orbit, in red for a representative high inclination orbit; the zones in black are excluded by the resonant eKBOs. A range of at least several degrees near each of these representative planes is allowed by our calculations.}  
\label{f:plotex}
\end{figure}

However, such geometric estimates of exclusion zones do not account for the planet's gravity. 
To take account of this,
we computed surfaces-of-section for the circular planar restricted three body problem, for several values of the planet mass 
in the range of $10^{-6}M_\odot$ to $10^{-3}M_\odot$, for the neighborhood of the 3/2, 5/2, 3/1 and 4/1 MMRs 
and the extreme eccentricities of the four eKBOs (0.85, 0.87, 0.85 and 0.7, respectively).  
Fig.~\ref{f:sos} shows a sample of these for planet mass $3\times10^{-5}M_\odot$ (approximately $10M_\oplus$).
From these surfaces-of-section, we determine the range of excluded longitudes of the planet by measuring the complement of the large 
libration zones of the 3/2 and 5/2 MMRs.  We also conservatively exclude 
a small range, $\pm5^\circ$, around the homoclinic points of the 3/1 and 4/1 resonant eKBOs. 
Fig.~\ref{f:plotex} shows our estimates of the excluded zones, projected on the sky, 
for a planet of mass $3\times10^{-5}M_\odot$; we plot a representative case of the planet's low inclination orbit (in green)
and also a representative case of its high inclination orbit (in red).  (Note that a range of orbital planes near each
of these representative cases is allowed by our calculations.)
We find that approximately half of the ecliptic longitudes are excluded in both cases; 
the excluded zone is not contiguous.

\section{Summary and Discussion}

We noticed that the four known longest period KBOs (Sedna (90377), 2010 GB174, 2004 VN112, 2012 VP113) 
could all be in mean motion resonances with an unseen distant planet.  The set of resonances is not uniquely determined.
In the present work we investigated the case of 
a hypothetical planet of orbit period $P'\simeq$~17,117~years (semimajor axis 
$a'\simeq665$~AU), which would have dynamically significant N/1 and N/2 period ratios with these four eKBOs.  
We give a broad sketch of how resonant geometries and resonant dynamics of these high eccentricity eKBOs 
help narrow the range of planet parameters; many details in this analysis would benefit from deeper investigation in future work and potentially improve planet parameters further. 

Our calculations suggest two possibilities for the planet's orbit plane: 
a plane moderately close to the ecliptic and near the mean plane of the 
four eKBOs ($i\simeq 18^\circ$, $\Omega\simeq 101^\circ$), 
or an inclined plane near $i\simeq 48^\circ$, $\Omega\simeq -5^\circ$.
The former offers dynamical stability of the four eKBOs by means of libration of the critical resonant angles, 
and the latter offers enhanced dynamical stability by means of additional libration of 
the argument of perihelion associated with {\it periodic orbits of the third kind} (a class of periodic orbits 
of the three dimensional restricted three body problem).  

In the case of coplanar MMRs, the planet's orbital eccentricity is limited 
to $e'\lesssim0.4$ by the putative 3/2 resonant orbit of Sedna, 
and to $e'\lesssim0.18$ by the putative 5/2 resonant orbit of 2010 GB174. 
For non-coplanar orbits, including the periodic orbits of the third kind,
the planet's eccentricity may be larger than these limits.
Preliminary numerical analysis finds that for relative inclinations up to $15^\circ$,
stable resonant librations are possible but uncommon for $e'\gtrsim0.3$.

A planet mass of $\sim10M_\oplus$ defines resonance widths that are similar to or exceed
the semimajor axis uncertainties of the observed eKBOs.  
Resonant libration periods would be on the order of a few megayears, 
much shorter than the age of the solar system.  
In the case of the inclined planet orbit, a planet of $\gtrsim10M_\oplus$ also supports librations about the
periodic orbits of the third kind, with libration periods shorter than the age of the solar system.
Both these estimates support the conjecture that the period ratio coincidences are of dynamical significance.

We determined exclusion zones of the current location of the planet in its orbital path 
(Fig.~\ref{f:plotex}), based on the MMRs of all four eKBOs;
these exclude just over half of the orbital path, assuming a $10M_\oplus$ planet. 

Our analysis supports the distant planet hypothesis, but should not be considered definitive proof of its existence.  
The orbital period ratios have significant uncertainties, so the near-coincidences with MMRs
may simply be by chance for a small number of bodies.  
We encourage further observations of these eKBOs to reduce the uncertainties in their orbital periods;
a factor of $\sim3$ reduction in the uncertainties would provide a clear test of our specific MMR hypothesis
for a $\sim10M_\oplus$ planet.  Future work should also examine other possible sets of MMRs.

As a final point, we note that the long orbital timescales in this region of the 
outer solar system may allow formally unstable orbits to persist for very long times, 
possibly even to the age of the solar system, depending on the planet mass; 
if so, this would weaken the argument for a resonant planet orbit. 
In future work it would be useful to investigate scattering efficiency as a function of the planet mass, 
as well as dynamical lifetimes of non-resonant planet-crossing
orbits in this region of the outer system.  
Nevertheless, the possibility that resonant orbital relations could be a useful aid to prediction and
discovery of additional high mass planets in the distant solar system makes a stimulating case for 
renewed study of aspects of solar system dynamics, such as resonant dynamics in the high eccentricity regime,
which have hitherto garnered insufficient attention. 

\acknowledgements
\section*{\it Acknowledgements}
RM thanks Scott Tremaine, Daniel Fabrycky, Sean Mills and Anthony Dobrovolskis for helpful comments on this work, 
Eric Christensen for an observer's perspective, and David Frenkel for many clarifying discussions.
RM and KV acknowledge funding from NASA (grant NNX14AG93G).
XW acknowledges funding from National Basic Research Program of China (973 Program) (2012CB720000) and 
China Scholarship Council.
This research made use of the NASA Astrophysics Data System Bibliographic Services.

\bibliographystyle{apj}

\end{document}